\documentclass[fleqn,usenatbib]{mnras}

\usepackage[T1]{fontenc}

\DeclareRobustCommand{\VAN}[3]{#2}
\let\VANthebibliography\thebibliography
\def\thebibliography{\DeclareRobustCommand{\VAN}[3]{##3}\VANthebibliography}

\usepackage{graphicx}	
\usepackage{amsmath}	
\usepackage{amssymb}	

\title[]{Post-merger evolution of double helium white dwarfs and distribution of helium-rich hot subdwarfs}

\author[]{
Jinlong Yu,$^{1}$\thanks{E-mail: yjl199502@163.com }
Xianfei Zhang,$^{2}$\thanks{E-mail: zxf@bnu.edu.cn }
 Guoliang L\"{u}$^{1}$\thanks{E-mail:  guolianglv@xao.ac.cn }
\\
$^{1}$School of Physical Science and Technology, Xinjiang University, Urumqi, 830046, China\\
$^{2}$Department of Astronomy, Beijing Normal University, Beijing 100875, China
}

\date{Accepted XXX. Received YYY; in original form ZZZ}

\pubyear{2015}

\begin{document}
\label{firstpage}
\pagerange{\pageref{firstpage}--\pageref{lastpage}}
\maketitle

\begin{abstract}
The mergers of double helium white dwarfs are believed to form isolated helium-rich hot subdwarfs. Observation shows that the helium-rich hot subdwarfs can be divided into two subgroups based on whether the surface is carbon-rich or carbon-normal. But it is not clear whether this distribution directly comes from binary evolution. We adopt the binary population synthesis (BPS) to obtain the population of single helium-rich hot subdwarfs according to the channel of double helium white dwarfs merger. We find that the merger channel can represent the two subgroups in the $T_{\rm{eff}}-\log g$ plane related to different masses of progenitor helium white dwarfs. For $Z$ = 0.02, the birth rates and local density of helium-rich hot subdwarf stars by the mergers of two helium white dwarfs is $\sim 4.82 \times 10^{-3}$ $\rm yr^{-1}$ and $\sim$ 290.0 $\rm kpc^{-3}$ at 13.7 Gyr in our Galaxy, respectively. The proportion of carbon-rich and carbon-normal helium-rich hot subdwarfs are 32$\%$ and 68$\%$, respectively.
\end{abstract}

\begin{keywords}
binaries: close --- stars: fundamental parameters --- stars: white dwarfs
\end{keywords}



\section{Introduction}

 In general, over 95 percent of stars will end their lives as white dwarf stars (WDs) and take a long time to cool down; thus, they are abundant and long-lived. More than half of the stars are in binary systems \citep{Han2020}. Therefore, after the interaction of binary stars, we can expect that many binary star systems contain at least one white dwarf, or even both of them are white dwarfs. According to different chemical compositions, white dwarfs can be divided into helium-white dwarfs (HeWDs), carbon-oxygen white dwarfs (CO WDs), and oxygen-neon-magnesium white dwarfs (ONeMg WDs).  Almost all the helium WDs result from binary star interaction because the evolution timescale for a single star to evolve into a HeWD would exceed Universe's age by far \citep{Bergeron1992,Han1998}. According to theory and observation, there are quite a few double HeWDs in the Galaxy, and they are also important sources of gravitational waves (GW) \citep{Bours2014}. Some of the close double HeWDs will be merged within a Hubble time due to orbital decay by GW radiation. The merger of two HeWDs provides a natural mechanism for producing extremely He-rich stars, e.g., He-rich hot subdwarfs. This channel is considered to be one of the vital scenarios to explain the formation of single He-rich hot subdwarfs \citep{Webbink1984,Saio2000,Han2002,Han2003,Zhang2012}.
 Also, this merger channel are supported by the observation distribution of He-rich hot subdwarfs in the Galactic \citep{Luo2019,Luo2020}.

 According to the surface effective temperature and spectra, hot subdwarfs are traditionally grouped into subdwarf B (sdB), subdwarf OB (sdOB), and subdwarf O (sdO). The classes sdOB and sdB are often merged into sdB owning to their sdB-like spectra with weak He II \citep{Heber2009,Heber2016}. Some hot subdwarfs have He-dominated spectra, which are known as He-rich hot subdwarfs. According to the recently complete samples of hot subdwarfs in Gaia, about $10\%$ of the hot subdwarf stars are He-rich hot subdwarfs \citep{Geier2017,Geier2019,Geier2020}. He-rich hot subdwarfs are subdivided into carbon-rich and nitrogen-rich (carbon-normal) according to their surface abundance of carbon and nitrogen\citep{Jeffery1997,Drilling2003,Stroeer2007,Jeffery2021}.
 By analyzing the spectrum, \cite{Hirsch2009} obtain carbon abundances (log $\beta_{C}$) and nitrogen abundances (log $\beta_{N}$) that are given by mass fraction, i.e.,$\rm log\beta_{i}=\frac{m_i}{m_H+m_{He}+m_c+m_N}$. They reported 16 carbon-rich hot subdwarf stars with surface abundances $-$2.60 $\leq$ log $\beta_{C}$ $\leq$ $-$1.56 and 17 carbon-normal hot subdwarf stars with surface abundances $-$4.52 $\leq$ log $\beta_{C}$ $\leq$ $-$4.14. Besides, the nitrogen abundances of carbon-rich objects are in the range $-$3.60 $\leq$ log $\beta_{N}$ $\leq$ $-$2.29, and the nitrogen abundances of carbon-normal stars are in the range $-$2.81 $\leq$ log $\beta_{N}$ $\leq$ $-$2.20.  \cite{Naslim2010} obtain a carbon-rich single hot subdwarf stars with surface abundances log $\beta_{C}$ $=$ $-$2.46 and four carbon-normal hot subdwarf stars with surface abundances $-$4.34 $\leq$ log $\beta_{C}$ $\leq$ $-$3.56. Following \citet{Hirsch2009}, similarly, we define log $\beta_{C}> -2.63$
 for C-rich He-rich hot subdwarfs, and log $\beta_{C}\leq -2.63$ for C-normal He-rich hot subdwarfs. The values of $-$2.63 corresponed to solar surface carbon abundance.

  \begin{figure}
	\includegraphics[width=\columnwidth]{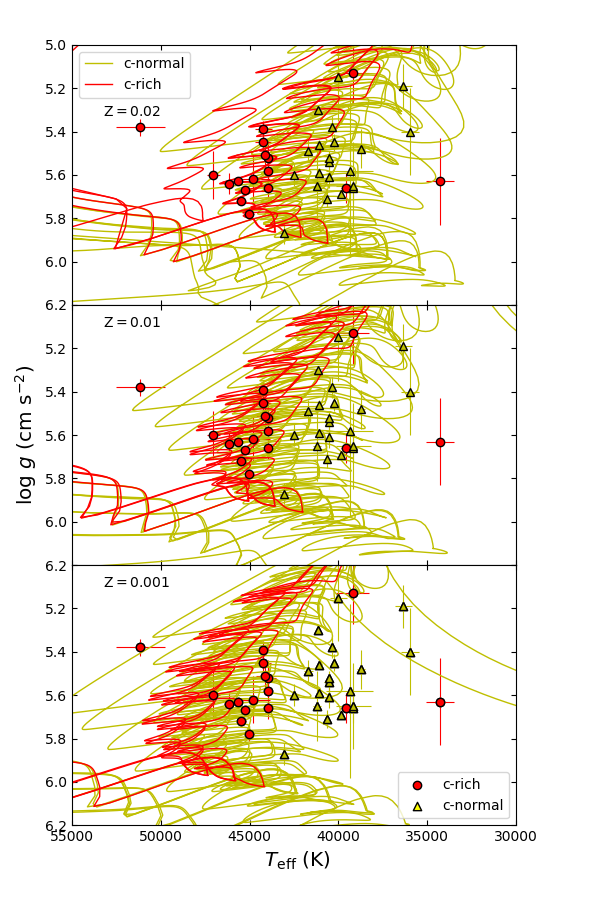}
    \caption{Evolutionary tracks of different double HeWDs merger remnants in the Teff - log g plane with different metallicity, e.g.,  Z= 0.02, 0.01, and 0.001. The red and yellow lines represent C-rich mergers and C-normal mergers, respectively. The red dots and yellow triangles indicate the C-rich and C-normal hot subdwarfs, which are observed by \citet{Hirsch2009} and \citet{Naslim2010}.}
    \label{fig1}
\end{figure}
\begin{figure}
	\includegraphics[width=\columnwidth]{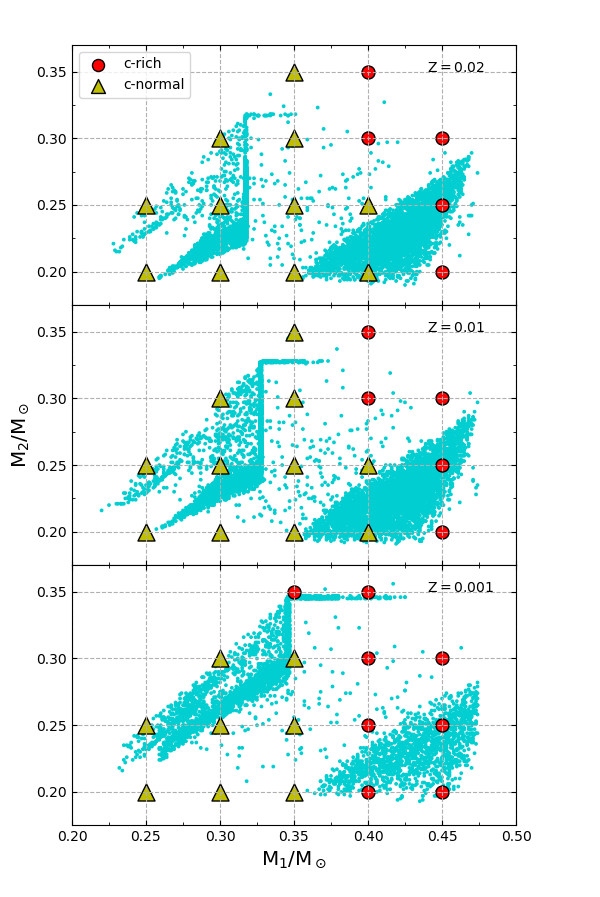}
    \caption{Models of double HeWDs merger remnants on the MWD-MWD planes. Each panel shows the models with different metallicity, e.g.,  Z= 0.02, 0.01, and 0.001, respectively. The red circles indicate C-rich mergers, and the yellow triangles indicate C-normal mergers.  The dots on the background show the masses of double HeWD obtained in the binary population synthesis.}
    \label{fig2}
\end{figure}

 In addition to the different abundances of elements, observations show that on the $T_{\rm eff}-logg$ diagram, the two types of hot subdwarfs, C-rich and C-normal, are obviously in two different temperature regions, and the dividing line is roughly 43000 K.
 An essential question is whether there is a real gap between C-rich and C-normal He-rich hot subdwarfs on the $T_{\rm eff}-logg$ diagram. The observational  bias effects (e. g., lack of statistically significant sample, errors of effective temperature, etc.) can make the gap, and theoretical evolutionary channel also can directly form the gap.
 To represent the mergers of double HeWDs, \cite{Zhang2012} build models that combine fast and slow accretion processes and their models are in agreement with the observation of surface effective temperatures, surface gravities, and abundances (in particular C and N). They show that the C-rich and C-normal hot subdwarfs can form from the mergers of different masses of progenitor HeWDs, e.g., the C-rich stars produced by high-mass HeWDs and C-normal stars made by low-mass HeWDs, respectively. However, it's difficult to explain why these two subgroups exist due to only a few models calculated in their work.

 In this letter, from the perspective of binary star population synthesis (BPS), we have extended the work of \cite{Zhang2012}. The results show that the distribution of C-rich and C-normal hot subdwarfs is probably a direct result of the helium white dwarfs' merger channel. Section \S 2 describes the stellar models and the simulation method of BPS. Results calculated from BPS and conclusions are presented in Sections \S 3 and \S 4, respectively.

\section{MODEL}

\subsection{Merger remnants}
We calculate the merger remnants by taking advantage of the detailed stellar evolution code Modules for Experiments in Stellar Astrophysics (\texttt{MESA}, version 12115, \citealt{paxton2011, paxton2013, paxton2015}). Following the methods of \citet{Zhang2012} and \citet{Zhang2014}, we generate the HeWDs models by evolving a 1.1 $\rm M_{\odot}$ zero-age main-sequence star (ZAMS) with metallicity Z= 0.02, 0.01, and 0,001,  then remove the complete hydrogen envelope by a high mass-loss rate until the He core reach the required mass. The naked He core evolves into the WD cooling phase straightly. Then we obtain the HeWD models when the logarithmic surface luminosity log(L /$L_{\odot}$) = $-$2. The masses of HeWDs range from 0.2 $\rm M_{\odot}$ to 0.45 $\rm M_{\odot}$ in steps of 0.05 $\rm M_{\odot}$ in our calculation.

\begin{figure*}
\centering
\begin{tabular}{lr}
\includegraphics[totalheight=3.5in,width=5.25in,angle=0]{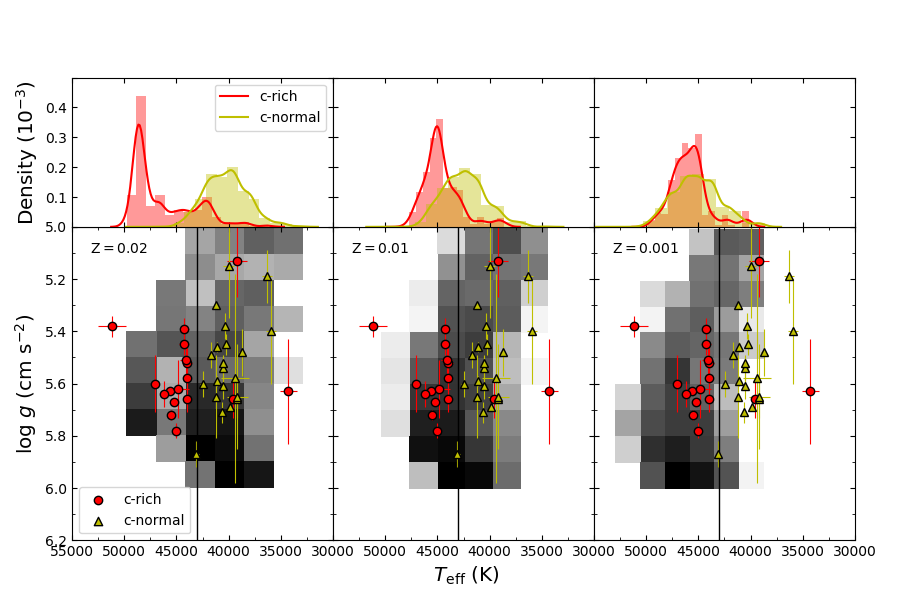}
\end{tabular}
    \caption{Distribution of the He-rich hot subdwarfs formed via the double HeWDs merger channel. The red dots and yellow triangles represent the C-rich and C-normal hot subdwarfs, respectively. The observed samples are from \citet{Hirsch2009} and \citet{Naslim2010}. In the top three panels, red and yellow lines represent the C-rich and C-normal normalized number density in the effective temperature range, respectively. The background grayscale distribution represents the relative number distribution of merged white dwarfs for He-rich hot subdwarfs.}
    \label{fig3}
\end{figure*}

Due to the SPH simulation of \cite{Dan2014} and \citet{Zhang2012} calculation, once the secondary (low-mass HeWD) are disrupted, the rest merger process can decompose into two critical accretion processes before and after. (1) a hot and fully convection envelope surrounds the primary formed by fast accretion rate,  which contains about half of the secondary's masses; (2) a disk created by the rest masses of the secondary, and then quickly disrupted by expanding envelope, which are still in a relative fast accretion \citep{Zhang2014}. Considering the convergence, we assume two similar steps to represent such processes.  We use a $10^{-3}$ $\rm M_{\odot}$ $\rm yr^{-1}$ accretion rate for half of the secondary's masses to create a hot envelope and make it thoroughly mix, then transfer the rest masses to the primary by the same accretion rate to represent the disk disruption process. It should be pointed out that the accretion rate we used is not consistent with the accretion rate given by the SPH simulation. We use this accretion rate value mainly to reproduce the similar temperature profile generated in the simulation. After the accretion processes, we obtain a series of post-merger models with different masses. Then, we evolve the post-merger models into the WD phase again.

\subsection{Evolutionary tracks of post-merger}

Figure \ref{fig1} shows the models with different metallicity, e.g., Z = 0.02, 0.01, and 0.001, respectively. As shown by \cite{Zhang2012}, post-mergers will undergo a series of inward shell helium flash processes, which will last for about 0.6 - 2.5 million years. When the shell helium flash ends, the central helium starts to burn and enters the hot subdwarf star's main sequence phase. For post-mergers with different masses, they have different surface carbon abundances. Generally, the greater the mass, the higher the abundance of carbon on the surface. As mentioned earlier, we define hot subdwarfs which surface carbon abundance exceeds the solar abundance as C-rich type, and vice versa as C-normal type. The evolutionary tracks in red and yellow represent the C-rich and C-normal post-mergers, respectively. We compared the evolution tracks with samples of \citet{Hirsch2009} and \citet{Naslim2010}. Except for the evolution model with Z=0.001, the observed samples are in good agreement with our models.  The C-rich type is mainly located in the high-temperature area, and the C-normal type is located in a relatively low-temperature area. However, this distribution feature will also be affected by the mass distribution of the white dwarfs before merging; hence we need to perform further binary population synthesis calculations.

\subsection{Binary Population Synthesis}

We use BPS to investigate the mass distribution of progenitor double HeWDs. Following \cite{Lu2012}, \cite{Zhu2015}, \cite{Lu2017}, \cite{Zhu2017} and \cite{Wang2018}, we perform a Monte Carlo simulation of a population with $10^{7}$ primordial binaries. Starting from zero-age main-sequence (ZAMS), each binary system evolves using rapid binary star evolution code BSE \citep{Hurley2000, Hurley2002}. Finally, we select the pairs of double HeWDs which will merge within a Hubble time to obtain the whole double HeWDs populations.

 In our Monte Carlo simulation, we assumed that all stars are binaries and circular orbits. We adopt the initial mass function (IMF) of \citet{Miller1979} to generate the primary, and in the mass range of 0.08 to 100 $\rm M_{\odot}$. We obtain the secondary mass from a constant mass-ratio distribution \citep{Mazeh1992}, i.e. $n(q^{'})$ = 1, 0 $\leq$ $q^{'}$ $\leq$ 1, where $q^{'}$ = $M_{\rm d}$/$M_{\rm a}$. We use the distribution of orbital separations given by \citet{Han1998} for compact binaries:
\begin{equation}
	an(a)=
	\begin{cases}
	0.07(a/a_0)^{1.2},\qquad a\le a_0 \\
	0.07,\qquad \qquad a_0 \le a \le a_1,
	\end{cases}
	\label{eq:3}
\end{equation}
where $a_0=10\;\rm{R}_{\odot}$, $a_1=5.75\times 10^6\;\rm{R}_{\odot}$ = 0.13 $\rm pc$. According to the formula, this results in approximately 50 percent of stellar systems having an orbital period of more than 100 yr. These systems are considered single star systems here\citep{Chen2009, Lu2011,Lu2012}. We adopt the same value of the IMF parameters, the mass-ratio distribution, and the initial orbital separation distribution. Such parameters were used in some previous works, e.g., hot subdwarfs \citep{Han2002,Han2003,Zhang2012}; double WD mergers in the Galaxy \citep{Han1998,Zhang2014}; EL CVn stars \citep{Chen2017}, extremely low-mass WDs (ELM WDs) in DDs \citep{Li2019} and A-type subdwarfs \citep{Yu2019}. Figure \ref{fig2} shows the mass distribution of double white dwarfs obtained from BPS and our post-merger evolution models.

\section{Result }
Once we have the  evolutionary tracks of post-mergers and the masses distribution of progenitor double HeWDs, thus we can combine both results to represent the He-rich hot subdwarf distribution. About 3000 HeWD+HeWD binaries will undergo merger to produce hot subdwarf stars of the $10^{7}$ binary systems. Combining the mass distribution of HeWDs, Figure \ref{fig2} shows the types of hot subdwarfs generated by merging HeWDs of different masses and only the massive HeWDs can produce the C-rich hot subdwarfs.  Furthermore, our results show that to form C-rich hot subdwarfs, the total masses of the pre-mergers of double helium white dwarfs must be large enough, and the masses of the primary HeWDs must be large enough either.

 Figure \ref{fig3} shows the distribution of C-rich and C-normal He-rich hot subdwarf stars formed via the HeWDs mergers with different metallicity in the $T_{\rm{eff}}-\log g$ plane. The grayscale image shows the distribution of the number of hot subdwarfs formed by merging helium-white dwarfs. We also show the distribution of C-rich and C-normal subtypes in the temperature range. To compare with the observations, Figure \ref{fig3} shows the observation results of \cite{Hirsch2009} and \cite{Naslim2010}. We can see that C-rich and C-normal hot subdwarfs are in two different regions in the figure, roughly with 43000K as the dividing line. Our calculation results reproduce this distribution feature. It is worth pointing out that from the results of our helium white dwarf merger model, the distribution characteristics of C-rich and C-normal hot subdwarfs are not only affected by the masses of progenitors of HeWDs, but also by their metallicities. As the metallicity decreases, the distribution intervals of C-rich and C-normal gradually move closer. When the Z=0.001, the distributions of the two subtypes in the temperature range are almost mixed together, it isn't easy to distinguish.

\begin{figure}
  \centering
  \includegraphics[width=0.5\textwidth]{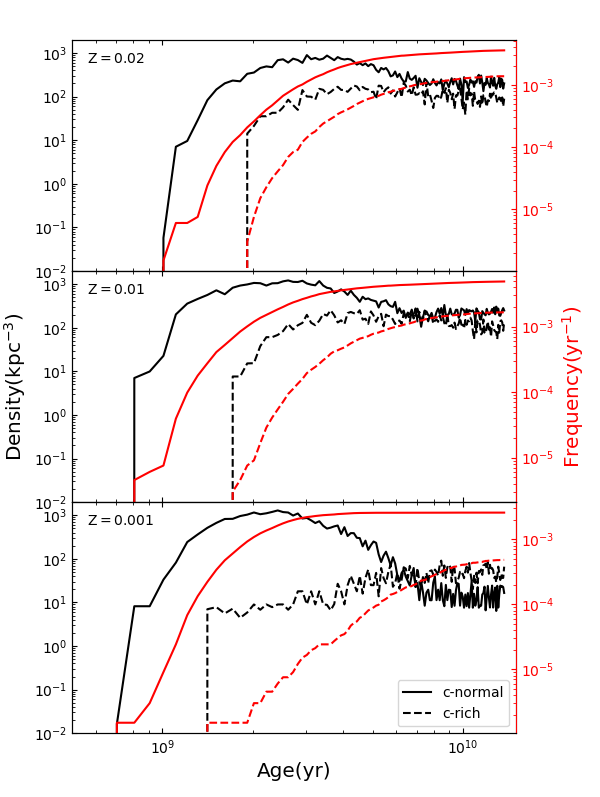}
  \caption{The local space density and birth rate of C-rich and C-normal He-rich hot subdwarf stars formed via the double HeWDs merger channel. The solid lines and dashed lines represent the C-normal and C-rich He-rich hot subdwarfs, respectively. }
  \label{fig4}
\end{figure}

  In \citet{Han1998}, the birth rates of the mergers of double HeWDs are $\sim 6.00 \times 10^{-3}$ $\rm yr^{-1}$ in the Galaxy. We obtain a similar merger rate of double HeWDs of $\sim 4.82 \times 10^{-3}$ $\rm yr^{-1}$. Figure \ref{fig4} shows the birth rate of C-rich and C-normal He-rich hot subdwarfs. We assume a constant star formation rate of  5 $\rm M_{\odot}$ $\rm yr^{-1}$ over the past 13.7 Gyr, and the galaxy's volume is 500 $\rm kpc^{3}$. By our calculation, the local space density and birthrate of He-rich hot subdwarfs are $\sim$ 290.0 $\rm kpc^{-3}$ and $\sim 4.82 \times 10^{-3}$ $\rm yr^{-1}$ at 13.7 Gyr for Z=0.02, respectively. Meanwhile, the proportion of C-rich and C-normal He-rich hot subdwarfs are 32$\%$ and 68$\%$, respectively. We expect that the number of C-rich and C-normal He-rich hot subdwarfs we may observe in the Milky Way are $\sim 4.64 \times 10^{4}$ and $\sim 9.86 \times 10^{4}$, respectively. For the metallicity Z=0.01 and 0.001, the birthrates are $\sim 6.23 \times 10^{-3}$ $\rm yr^{-1}$ and $\sim 2.80 \times 10^{-3}$ $\rm yr^{-1}$ at 13.7 Gyr; the local space densities are $\sim$ 379.0 $\rm kpc^{-3}$ and $\sim$ 83.0 $\rm kpc^{-3}$ at 13.7 Gyr.

\section{Discussion and CONCLUSIONS}

  It is generally believed that hot subdwarfs are a special kind of stars formed by the evolution of binary stars, and the details of their formation are still not exact enough. He-rich hot subdwarfs, which are almost entirely composed of helium, are a very special subtype. The merger of double helium white dwarfs is considered an essential channel for forming He-rich hot subdwarfs. Observedly, He-rich hot subdwarfs can be divided into two groups: C-rich and C-normal in terms of temperature distribution. It is of great significance for us to understand the formation of hot subdwarfs to clarify whether this distribution feature is naturally derived from binary stars' evolution or caused by other reasons, such as observation effects. In this paper, we constructed a series of double HeWD merge models following \citet{Zhang2012} and performed evolutionary calculations on post-mergers. Combined with binary population synthesis, we have studied the distribution characteristics of He-rich hot subdwarfs formed by merging channels.

  We found that the He-rich hot subdwarf formed by merging channels is distributed in two different regions on the $T_{\rm{eff}}-\log g$ plane. Its interval range is consistent with the corresponding C-rich and C-normal He-rich hot subdwarf observations. Meanwhile, we also obtain a similar surface abundance of carbon. The main reason for this difference is that the C-rich hot subdwarfs are derived from the merger of the larger-masses double helium white dwarf star than the C-normal hot subdwarfs. Furthermore, as the metal content increases from our calculation results, the separation between these two regions will become larger. Therefore, the observed distributions of C-rich and C-normal He-rich hot subdwarf in different temperature regions are most likely the results of binary star evolution. Of course, due to the limited number of samples, we look forward to more observations to confirm this.

  It is worth pointing out that the surface carbon abundance is directly related to the merging model's temperature profile and the element mixing process caused by convection. The merger's maximum temperature is related to the accretion rate used in the merger process simulation. The maximum accretion rate in our calculation can only reach $10^{-3}$ $\rm M_{\odot}$ $\rm yr^{-1}$, which is lower than the accretion rate used by \citet{Zhang2012}, so the surface carbon element is also lower than \citet{Zhang2012}. However, the distribution of C-rich and C-normal regions is mainly related to the mass of double helium white dwarfs. The accretion rate only affects the abundance of carbon. Thus the accretion rate is not affected much and does not affect our conclusion. In future work, we will further improve our model calculations and look forward to more accurately reproducing the formation and evolution of these He-rich hot subdwarfs. Besides, the other formation channel of He-rich hot subdwarfs, e.g., late hot flasher scenarios, are also worthy of consideration in future work.
  Although we can explain the distribution characteristic of helium-rich hot subdwarf by the evolutionary channel, we still cannot exclude the possibility of the lacking sufficiently large observational sample. Therefore, a much larger observational sample would be helpful to examine our results.

\section*{Acknowledgements}
This work received the generous support of the National Natural Science Foundation of China,
project Nos. U2031204, 11763007, 11863005, 12073006 and U2031203.
We would also like to express our gratitude to the Tianshan Youth Project of Xinjiang No. 2017Q014.

\section*{Data Availability}

No new data were generated or analysed in support of this research.



\bibliographystyle{mnras}
\bibliography{example} 





\bsp	
\label{lastpage}
\end{document}